\documentclass[pra,twocolumn,amsmath,amssymb,superscriptaddress,showpacs]{revtex4}
\usepackage{graphicx}% Include figure files
\usepackage{bm}
\usepackage{amsmath,amssymb}
\usepackage[usenames]{color}
\newcommand{\ii}{\mathbb{I}}

\newcommand{\bra}[1]{\langle #1 \vert}

\newcommand{\ket}[1]{\vert #1 \rangle}

\begin{document}
\title{Quantum proofs can be verified using only single qubit measurements}  
\author{Tomoyuki Morimae}
\email{morimae@gunma-u.ac.jp}
\affiliation{ASRLD Unit, Gunma University,
1-5-1 Tenjin-cho Kiryu-shi Gunma-ken, 376-0052, Japan}
\author{Daniel Nagaj}
\email{daniel.nagaj@savba.sk}
%\affiliation{Simons Institute for the Theory of Computing, 121 Calvin Lab \#2190, UC Berkeley, 94720-2190 Berkeley, CA, USA}
%\affiliation{Faculty of Physics, University of Vienna, Boltzmanngasse 5, 1090 Vienna, Austria}
\affiliation{Institute of Physics, Slovak Academy of Sciences, D\'{u}bravsk\'{a} cesta 9, 84511 Bratislava, Slovakia}
\author{Norbert Schuch}
\affiliation{JARA Institute for Quantum Information, RWTH Aachen
University, 52056 Aachen, Germany}

\date{\today}
            
\begin{abstract}
QMA (Quantum Merlin Arthur) is the class of problems which, though
potentially hard to solve, have a quantum solution which can be verified
efficiently using a quantum computer. It thus forms a natural quantum
version of the classical complexity class NP (and its probabilistic variant MA, Merlin-Arthur games), where the verifier has only classical computational resources. In this paper, we study what happens when we restrict the quantum resources of the verifier to the bare
minimum: individual measurements 
on single qubits received as they come, one-by-one.
We find that despite this grave restriction, 
it is still possible to soundly verify any problem in QMA
for the verifier with the minimum quantum resources possible,
without using any quantum memory or multiqubit operations. 
We provide two independent proofs
of this fact, based on measurement based quantum computation and the local
Hamiltonian problem, respectively. The former construction
also applies to QMA$_1$, i.e., QMA with one-sided error.
\end{abstract}

\pacs{03.67.-a, 03.67.Ac, 03.67.Lx}
\maketitle  
%------------------------------------

\section{Introduction}

One of the key questions in computational complexity is to determine the
resources required to find a solution to a certain problem. On the other
hand, even if we do not know how to produce a solution with some given
resources, we can still ask whether those resources allow us to verify
the correctness of a given solution. Most importantly, this gives rise to
the complexity class NP, the class of decision problems whose `yes'
instances have proofs that can be efficiently (in polynomial time, with
perfect soundness) verified by a deterministic classical computer.  This
concept can be generalized by allowing probabilistic verification of
proofs, leading to the class MA (Merlin-Arthur).

The natural quantum version of these classes is formed by the class
QMA (Quantum Merlin-Arthur)~\cite{Knill,Kitaev,WatrousQMA,Aharonov},
which consists of all problems whose `yes' instances have a ``quantum
proof'' (i.e., a quantum state) which can be efficiently verified in
polynomial time by a quantum computer.  The prototypical QMA-complete
problem is to determine the ground state energy of a Hamiltonian with
few-body interactions, known as the Local Hamiltonian problem~\cite{KKR03}
(for a review of recent progress on QMA-complete problems, see
\cite{Bookatz}). Here, the proof is the ground state itself, whose energy
can be efficiently estimated using a quantum computer.

All these classes can be understood in terms of a game between an
all-powerful prover Merlin and a rational verifier Arthur with limited resources, where
Merlin tries to prove some statement to Arthur by sending
him a classical or quantum proof, which Arthur then verifies using a
classical (NP, MA) or quantum (QMA) computer.  Adding quantum mechanics
opens new doors for Merlin to cheat, but on the other hand, performing a
quantum computation as a verification procedure gives more power to Arthur
as well. Indeed, since Arthur can always start by measuring in the
computational basis, effectively treating the proof as classical, QMA is at
least as powerful as NP and MA.  

However, will we tip this balance when we restrict Arthur's power, and
supply him only with restricted quantum resources rather than access to a
full quantum computer?  Most extremely, we could imagine that Arthur does
not have access to a quantum memory and can only perform single-qubit
measurements on a sequence of qubits sent by Merlin one-by-one, possibly
in a restricted basis. It seems likely that the class of problems
that could be soundly verified in such a setup is much smaller than QMA. 

In this paper, we prove that this is not the case: Even if Arthur is
limited to single qubit measurements which are performed one-by-one on
qubits sent sequentially by Merlin, the class of problems which can be
proven this way still equals QMA.  We will show this in two distinct ways:
The first proof utilizes measurement-based quantum computing
(MBQC)~\cite{MBQC}, while the second proof is based on a scheme for
single-qubit measurement verification of the ground state for the Local
Hamiltonian problem~\cite{Kitaev}.  

Our first proof uses MBQC, a universal model of quantum computing where
adaptive single-qubit measurements are applied to a certain highly
entangled many-qubit state, such as the graph state~\cite{MBQC}.  In our
protocol, an honest Merlin sends Arthur the graph state coupled with a
witness state, qubit-by-qubit. As Arthur receives each qubit, he measures
it. Using such adaptive single-qubit measurements, 
he applies the desired QMA verification circuit via a MBQC scheme. On the other hand, if
Merlin is malicious, he does not necessarily send the correct graph state.
However, Arthur can verify whether he received the proper graph state by
measuring a list of stabilizers, which can again be done using only
single-qubit measurements, maintaining soundness of the protocol.  Our
proof relies on the idea of Ref.~\cite{Matt}, where graph state
verification is the basis of a multiprover quantum interactive proof system for BQP with a classical verifier and several entangled but non-communicating provers. Our usage of graph state verification is much simpler  because
Arthur does the measurements by himself, and therefore he does not 
need to test device independence, while his measurement choices are naturally hidden from Merlin. 
Interestingly, this MBQC approach also allows to reduce the required {\it classical} 
computational ability of Arthur to only XOR operations,
since it is known~\cite{Barz} that single-qubit unitary operations on a 
known single-qubit state, single-qubit measurements, and the classical XOR
gate are enough for classical universal computing.

In the second proof, instead of implementing a quantum verification
circuit, we look at the possibility of determining the ground state energy
of a local Hamiltonian using only single-qubit measurements. The trick is
to ask Merlin for the ground state, but to keep secret which of the terms
in the Hamiltonian we will measure. In fact, we show how to randomly
choose and perform a sequence of single-qubit measurements so that the
protocol remains complete and sound.  Our single-qubit measurement
approach to verifying ground states of a local Hamiltonian was first
presented by one of the authors in a Stack-exchange post
\cite{stackexchange} in response to a question by Lior Eldar. Note that
another protocol for this task can be derived from the recent multiprover
verification scheme by Ji \cite{JiVerification}.

The efficiency of our local-Hamiltonian based protocol is similar (up to
polynomial factors) to that of the MBQC approach, which is natural for
quantum circuits. However, the MBQC argument also applies to the
verification procedure of the complexity class QMA$_1$, i.e., QMA with
perfect completeness, while the Hamiltonian approach does not work there,
as it is inherently probabilistic even for an honest Merlin. On the other
hand, the local-Hamiltonian approach requires only Pauli measurements,
while the MBQC approach involves measurements outside the Clifford basis.

Note that recently, another direction of restricting Arthur's power -- to Clifford gate operations only -- has been considered
~\cite{CliffordQMA}.  
It was shown there that even if Arthur's power is
restricted to Clifford gate operations, the class equals to QMA. The basic tool there
is the universality of Clifford gates plus magic
states~\cite{magic}: Merlin sends Arthur many copies of magic states in
addition to the witness, and Arthur, who can perform only Clifford gate
operations, uses them for universal quantum computing. Even if a malicious
Merlin sends some other states pretending to be magic states, Arthur can
filter them to guarantee the soundness.  The result of this paper, showing
that restricting Arthur's power to single-qubit measurements does not
weaken the complexity class QMA (or ${\rm QMA}_1$), is therefore a step in a
similar direction.

Furthermore, Ref.~\cite{stoqMA} introduced the class stoqMA.
It is another variant of QMA, restricting Arthur's ability to
classical reversible gates on initial qubits prepared in
$|0\rangle$ or $|+\rangle$, and finally measurement of the output qubit in
the $X$ basis.
Unlike our case, stoqMA is not known to be equal to QMA; we know it is
contained in SBP.
On the other hand, what if we put restrictions on Merlin instead of Arthur?
Ref.~\cite{subset} has shown that even if the quantum witness
is restricted to be an equal-weight positive-constants superposition
of computational basis states, the class of problems provable in this way 
is still equal to QMA.

%%%%%%%%%%%%%%%%%%%%%%%%%%%%%%%%%%%%%%%%

\section{Preliminaries}
\subsection{QMA and its Verification Protocol}

Consider a language $L$ (i.e., the set of `yes' instances of a problem
such as Local Hamiltonian) and denote its instances by $x$, and the length
of the bit-string $x$ by $|x|$.  The language $L$ belongs to the class
QMA$(a,b)$ with $a-b\ge1/\textrm{poly}(|x|)$ if for each $x$, there
exists a polynomial-size quantum circuit $Q_x$ (from a uniform family of
circuits), working on $n=\textrm{poly}(|x|)$ qubits
and $m=\textrm{poly}(|x|)$ ancilla qubits
such that
\begin{itemize}
\item[1.]
({\it Completeness})
if $x\in L$, there exists an $n$-qubit witness state 
$|\xi_x\rangle$,
such that the result of the computational-basis measurement
on the first qubit of $Q_x\left(|\xi_x\rangle\otimes\ket{+}^{\otimes m}\right)$
is 1 with probability $\ge a$,
\item[2.]
({\it Soundness})
if $x\notin L$, 
the result of the computational-basis measurement 
on the first qubit of $Q_x\left(|\psi\rangle\otimes\ket{+}^{\otimes m}\right)$
is 1 with probability $\le b$
for any $n$-qubit input state $|\psi\rangle$.
\end{itemize}
Usually, each ancilla qubit is initialized in the state $|0\rangle$.
Here we choose the basis state $\ket{+}\equiv(\ket{0}+|1\rangle)/\sqrt{2}$ instead, to make the procedure compatible with the standard notation for MBQC.

A priori, the class QMA$(a,b)$ depends on the completeness and soundness
parameters $a$ and $b$. 
%However, it can be shown that QMA$(a,b)$ is in fact independent of $a$ and $b$, as long as $a-b\ge1/\mathrm{poly}(|x|)$, $a\le 1-e^{-O(|x|)}$, $b\ge e^{-O(|x|)}$~\cite{Aharonov}; this class is then simply called QMA. 
%\note{One more try (in blue). [D]}
However, there are several ways to amplify \cite{NWZamplification} the parameter $a$ to make it close to 1 and the parameter $b$ close to 0. In fact, as long as $a-b\ge1/\mathrm{poly}(|x|)$, 
and $a\le 1-e^{-\textrm{poly}(|x|)}$, $b\ge e^{-\textrm{poly}(|x|)}$,
the power of this protocol does not change~\cite{Aharonov}; 
we thus simply call QMA$(a,b)$ with these restrictions QMA. 

%This means $a$ can be amplified to a number exponentially close to $1$ and $b$ can be transformed to one exponentially close to zero, without changing the power of the class.
On the other hand, the case $a=1$ is special,
in that it requires the existence of a proof which is accepted with unit
probability (perfect completeness); in that case, with the conditions on $b$ as before, the class is denoted by $\textrm{QMA}_1$.

\subsection{Measurement based quantum computing}

Measurement based quantum computing (MBQC) is a model of universal quantum
computing proposed by Raussendorf and Briegel~\cite{MBQC}.  In this model,
universal quantum computation can be realized by the preparation of a
certain many-qubit {\em resource state}, followed by sequential adaptive
single-qubit local measurements on the resource state's qubits. The {\em
cluster state} (or the graph state)~\cite{MBQC} is the canonical example
of such a resource state.  Let $G=(V,E)$ be a graph, where $V$ is the set
of vertices and $E$ is the set of edges.  The cluster state $\ket{G}$ on
the graph $G$ is defined by 
\begin{align*} 
	\ket{G} \equiv
	\left(\bigotimes_{e\in E}CZ_e\right) \ket{+}^{\otimes |V|},
\end{align*} 
where a qubit in the state $\ket{+}$ is located on each
vertex of $G$, and $CZ_e$ is the controlled-$Z$ gate on the edge $e$.  It
is known that, for example, the graph state on the two-dimensional square
lattice is a universal resource state~\cite{MBQC}.

The cluster state can also be used to apply a quantum circuit to an
arbitrary input state $\ket\psi$. To this end, one splits the
vertices into two sets $V_1$ and $V_2$. The vertices in $V_1$ are prepared
in $\ket{+}^{\otimes |V_1|}$, while the vertices in $V_2$ are prepared in
$\ket\psi$, and subsequently, $\otimes_{e\in E}CZ_e$ is applied along the
edges of the graph as shown in Fig.~1 (i.e., a square lattice on $V_1$
connected to the vertices in $V_2$). Starting from a state of this form,
one can then carry out any desired quantum computation on $\ket\psi$ by
doing single-qubit measurements only.

\begin{figure}[htbp]
\begin{center}
	\includegraphics[width=0.26\textwidth]{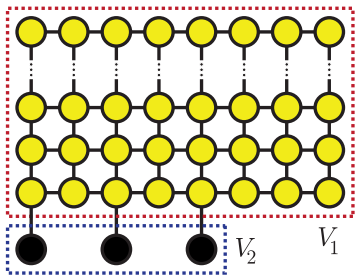}
\end{center}
\caption{
The graph $G=(V,E)$. Arthur expects Merlin to send him a universal graph state on vertices $V_1$ and a witness state on vertices $V_2$, coupled together by controlled-Z gates across the edges between $V_1$ and $V_2$. 
}
\label{stabilizer_fig}
\end{figure}

\section{MBQC approach}

Let us now give our first result -- a QMA verifier which uses only
sequential single-qubit measurements based on MBQC. The basic idea is as
follows: Given a QMA problem specified by a verifier circuit $Q_x$, Arthur
asks Merlin to send him the witness state $\ket\psi$, coupled to a graph
state, which allows Arthur to implement $Q_x$ on $\ket\psi$ using MBQC, as
in Fig.~\ref{stabilizer_fig}.  With an honest Merlin, Arthur can run
$Q_x$ on $\ket\psi$ using MBQC to verify the proof; on the other hand, we
will show that a cheating Merlin can be caught by testing the graph state
using only single-qubit measurements.  Note that there is no need for
Arthur to explicitly ask for ancilla qubits, as these are implicit in a
sufficiently large graph state -- we can take the ancillas required for
computation to be some of the $\ket{+}$ states in the graph state (they
are also coupled together with the rest of the graph state by
controlled-$Z$ gates).  

Let us thus consider a graph $G=(V,E)$ as in Fig.~\ref{stabilizer_fig}.
We denote the set of vertices in the red (blue) region by $V_1$ ($V_2$),
and define $N\equiv|V_1|=\mathrm{poly}(|x|)$.  Note that $|V_2|=n$.  We also denote
the set of edges in the red region by $E_1$.  Let $E_{\textrm{conn}}$ be
the set of edges that connect the red region and blue region, i.e.,
$E_{\textrm{conn}}=E-E_1$.

Now consider the following interactive proof scheme.  Merlin sends a state
$\rho$ on $G$ to Arthur, qubit-by-qubit, in a predefined order.  If Merlin
is honest, $\rho$ is the graph state plus witness as described above, but
if Merlin is malicious, $\rho$ can be any state.  Arthur then runs the
following protocol.  With some probability $q$, which will be specified
later, Arthur uses $\rho$ to run the verifier circuit $Q_x$ on $\ket\psi$
using MBQC.  If the computation accepts (rejects), Arthur accepts
(rejects).  On the other hand, with probability $1-q$, Arthur performs the
following {\em stabilizer test}: He randomly generates an $N$-bit string
$k\equiv(k_1,...,k_N)\in\{0,1\}^N$, and measures the operator 
\begin{align*}
	s_k\equiv\prod_{j\in V_1}g_j^{k_j},
\end{align*}
(note that we choose $j$ only from the vertices $V_1$, and not on the vertices $V_2$ where the witness is located), where the operator 
\begin{align*}
	g_j\equiv X_j\bigotimes_{i\in S_j}Z_i
\end{align*}
is a stabilizer of the graph state which applies $X$ to vertex $j$, and
$Z$ to its neighbors (here, $S_j$ denotes the set of the nearest-neighbor
vertices of vertex $j$).
Since $s_k$ is a tensor product of $X$, $Z$, and $Y$, Arthur can
(destructively) measure
$s_k$ by measuring each qubit independently, as it arrives from Merlin,
without the need for memory or multi-qubit operations.  If the result is
$+1$ ($-1$), 
the test passes (fails), and Arthur accepts (rejects) Merlin's proof.
The probability of passing the stabilizer test is
\begin{align*}
p_{\textrm{pass}}=\frac{1}{2^N}\sum_{k\in\{0,1\}^N}
\mbox{Tr}\left(\frac{\ii+s_k}{2}\,\rho\right).
\end{align*}

We will now show that this protocol is complete and sound.
First let us consider the case of $x\in L$.
Since Merlin is honest, he sends Arthur
\begin{align*}
	\left(\bigotimes_{e\in E_{\textrm{conn}}}CZ_e\right)
	\left(|G\rangle_{V_1}\otimes\ket{\xi_x}_{V_2}\right),
\end{align*}
where $|G\rangle$ is the graph state on the graph $(V_1,E_1)$, 
and $\ket{\xi_x}$ is the correct witness state on $V_2$,
which is accepted with probability $\ge a$ in the original QMA
protocol.
The probability of passing the stabilizer test is 1, since 
we are measuring the stabilizers of a proper graph state. 
On the other hand, when Arthur chooses to do the computation, 
he will accept Merlin's proof 
with probability larger than $a$ (the completeness of the QMA protocol). 
Therefore, the overall 
acceptance probability is
\begin{align}
p_{\textrm{acc}}^{x\in L}\ge qa+(1-q)\equiv\alpha. \label{accHONEST}
\end{align}

Next, let us consider the case $x\notin L$.  If Merlin wants to fool
Arthur, he has two options: Either, he sends a state that is close to the
correct proof (graph state plus witness) and thus has a high probability 
$p_{\textrm{pass}}$ of passing the stabilizer test; 
as we will show, such a state will fail the
QMA verification. Otherwise, Merlin could try to send a state which is
farther away from the correct proof and thus can pass the QMA
verification, but such a state will fail the stabilizer test.

Let us thus fix some (small) $\epsilon>0$ (which will be specified later),
and first consider the case where Merlin sends Arthur a state with $p_{\textrm{pass}}\ge1-\epsilon$.  Then, 
\begin{align*}
	\mbox{Tr}\left(
	\prod_{j\in V_1}\frac{\ii+g_j}{2} \,\rho
	\right)
	=
	\mbox{Tr}\left(
		\frac{1}{2^N}\sum_{k\in\{0,1\}^N} s_k \,\rho
	\right) 
	\ge 1-2\epsilon,
\end{align*}
using the relation
\begin{align*}
\prod_{j\in V_1}\frac{\ii+g_j}{2}
=\frac{1}{2^N}\sum_{k\in\{0,1\}^N}s_k.
\end{align*}
Let $W\equiv\bigotimes_{e\in E}CZ_e$ be the product of controlled-Z gates 
on all edges.  Then,
\begin{align*}
	\mbox{Tr}\left(	\prod_{j\in V_1}\frac{\ii+g_j}{2} \,\rho\right)
	&=
	\mbox{Tr}\left(
			\prod_{j\in V_1}\left(W\frac{\ii+g_j}{2}W\right)
			W\rho W
		\right)\\
	&=
	\mbox{Tr}\left(
		\bigotimes_{j\in V_1}\frac{\ii+X_j}{2}
		W\rho W\right)\\
	&=
	\mbox{Tr}(
	\ket{+}\bra{+}^{\otimes N}_{V_1}\otimes \ii_{V_2}
W\rho W)\\
&=
	\mbox{Tr}\left(
		\ket{+}\bra{+}^{\otimes N}_{V_1}
		\mbox{Tr}_{V_2}\left(W\rho W\right)\right)\\
&= F\left(\ket{+}\bra{+}^{\otimes N}_{V_1},\mbox{Tr}_{V_2}\left(W\rho W\right)\right)^2\\
&= \max_{w_{V_2}}
	F\left(\ket{+}\bra{+}^{\otimes N}_{V_1}\otimes w_{V_2},W\rho W\right)^2\\
&= F\left(W\left(\ket{+}\bra{+}^{\otimes N}_{V_1} \otimes w_{V_2}^*\right) W,\rho \right)^2,
\end{align*}
where $\mbox{Tr}_{V_2}$ is the partial trace over
$V_2$, and $F(\rho,\sigma)\equiv\mbox{Tr}\sqrt{\sqrt{\rho}\,\sigma\sqrt{\rho}}$ 
is the fidelity between $\rho$ and $\sigma$. 
We have relied on the identity~\cite{HayashiMorimae}
\begin{align*}
	&F\left(|+\rangle\langle +|^{\otimes N}_{V_1},\mbox{Tr}_{V_2}(W\rho W)\right) \\
	&\qquad\qquad=
	\max_{w_{V_2}}F\left(|+\rangle\langle+|^{\otimes N}_{V_1}\otimes w_{V_2},W\rho W\right),
\end{align*}
and called $w_{V_2}^*$ the state that achieves the maximum.  By using the
relationship between the trace distance and fidelity, we obtain
\begin{align*}
	&\frac{1}{2}\left\|
			W\left(|+\rangle\langle +|^{\otimes N}_{V_1}
		\otimes w_{V_2}^*\right) W-\rho
		\right\|_1\\
	\le& \sqrt{1-F\left(W\left(
			|+\rangle\langle +|^{\otimes N}_{V_1}\otimes w_{V_2}^*
			\right) W,\rho\right)^2}\\
	\le& \sqrt{1-(1-2\epsilon)}
	= \sqrt{2\epsilon}.
\end{align*}
Therefore, the acceptance probability for a malicious Merlin that wants to pass the stabilizer test with probability greater than $1-\epsilon$ is bounded from above by 
\begin{align}
	p_{\textrm{acc},1}^{x\notin L}\le q\left(b+\sqrt{2\epsilon}\right)+(1-q)
		\equiv \beta_1.
	\label{accPASS}
\end{align}

On the other hand, what happens when Merlin sends Arthur a 
state that passes the stabilizer test with probability at most 
$p_{\textrm{pass}}<1-\epsilon$? 
In that case, we can assume that this false
state is far from the graph state, and that Merlin tweaked it in such a way that it 
passes the computational test with probability one. 
However,
the detection probability from the stabilizer test is enough to give us an
upper bound on the overall acceptance probability:
\begin{align}
	p_{\textrm{acc},2}^{x\notin L} < q+(1-q)(1-\epsilon)
		\equiv \beta_2. \label{accFAIL}
\end{align}

We now need to show that the acceptance probabilities
of Eq.~\eqref{accPASS} and Eq.~\eqref{accFAIL} are necessarily lower than
Eq.~\eqref{accHONEST}, resulting in an at least inverse-polynomial
completeness-soundness gap for the MBQC-based single-qubit measurement QMA
protocol. 

We will do this by finding a setting for $q$ and $\epsilon$ that gives us the best completeness-soundness gap. Let us look at the possible gaps:
\begin{align*}
	\Delta_1(q,\epsilon) &\equiv	\alpha-\beta_1 = qa-q(b+\sqrt{2\epsilon}),\\
	\Delta_2(q,\epsilon) &\equiv \alpha-\beta_2 = qa-q+\epsilon(1-q).
\end{align*}
It is optimal for Arthur to choose the value of $q$ that satisfies $\Delta_1=\Delta_2$:
\begin{align*}
	q^*=\frac{\epsilon}{1+\epsilon-b-\sqrt{2\epsilon}}.
\end{align*}
It is then straightforward to choose $a=\frac{2}{3}$, $b=\frac{1}{3}$ (using amplification for the original circuit), 
and $\epsilon=\frac{1}{2|x|^2}$, and obtain a completeness-soundness gap for the new MBQC-based protocol 
\begin{align*}
p_{\textrm{acc}}^{x\in L} - p_{\textrm{acc}}^{x\notin L} 
%= \Delta_1(q^*) = \Delta_2(q^*)
&\geq \Delta(q^*,\epsilon)
= \frac{\epsilon(a-b-\sqrt{2\epsilon})}{1+\epsilon-b-\sqrt{2\epsilon}}\\
&\ge \frac{\epsilon(a-b-\sqrt{2\epsilon})}{2}
= \frac{\frac{1}{3}-\frac{1}{|x|}}{4|x|^2}
%\\
%&\ge \frac{\frac{1}{12}}{4|x|^2}
%=
\ge \frac{1}{48|x|^2},
\end{align*}
for $|x|\geq 4$. We have thus proved the new protocol is complete and sound, with an inverse-polynomial promise gap.

Note that this also works for a perfectly complete original QMA protocol 
with $a=1$, since the honest acceptance probability remains perfect, $p_{\textrm{acc}}^{x\in L}=1$. Therefore,
this MBQC-based single-qubit measurement protocol is valid also for ${\rm QMA}_1$.

\section{Local Hamiltonian approach}

Let us now turn towards our second construction for a QMA verification
restricted to single qubit measurements.  Rather than implementing the
verifier circuit using MBQC, we now devise a way to perform a
restricted-quantum-power verification of proofs for a particular
QMA-complete problem: the Local Hamiltonian \cite{Kitaev}. %problem.  

The $k$-Local Hamiltonian promise problem asks whether a Hamiltonian
$H=\sum_{m=1}^M H_m$ made from $k$-local (i.e., $k$-body) terms has a
ground state energy below some $E_a$, or above some $E_b$, with a promise
gap
$E_b-E_a > 1/\textrm{poly}(|x|)$. This problem is QMA-complete in short
because a successful verification of a proof using a quantum circuit $Q_x$ can be
encoded into the ground state of a particular Hamiltonian, and measuring
the energy of a state is a simple task using a quantum computer.  
This can either be done by picking a term $H_m$ at random according to its
norm and measuring the expectation value of this term~\cite{Aharonov}
(this procedure can be amplified using multiple copies of the state), or
in one go by doing phase estimation of $e^{-iH\tau}$ \cite{LHphase}.  Yet, both 
these schemes require joint measurements on at least $k$ qubits. However,
as we will show in the following, it is possible to estimate 
$\bra\psi H\ket\psi$ with single-qubit Pauli measurements only.

The trick is to decompose the Hamiltonian terms in the Pauli basis (or
any other local basis), pick one of the Pauli terms according to a
particular probability distribution, and measure this term qubit-by-qubit.
Let us present the scheme in full detail. Consider an $N$-qubit system,
and a $k$-local Hamiltonian 
$
H=\sum_{m=1}^{M} H_m,
$
together with a promise pair $E_a, E_b$, separated by 
$E_b-E_a\geq 1/\textrm{poly}(|x|)$. 
Each of the terms $H_m$ is $k$-local, acting non-trivially on at most $k$ qubits. We can thus decompose it in the Pauli basis as
\begin{align*}
H_m = \sum_{S\in \mathcal{P}} c^m_S S,
%\label{Hm}
\end{align*}
where $\mathcal{P}$ is the set of tensor products of $N$ Pauli matrices (or
identities).  We can then rearrange the Hamiltonian $H$ as
\begin{align*}
H &= \sum_{S\in \mathcal{P}} 
\underbrace{\left(\sum_{m=1}^{M} c^m_S\right)}_{d_S} S, 
%\label{Horig}
\end{align*}
labeling $d_S$ the sum of all the prefactors 
that contribute to the particular $k$-local Pauli operator $S$.
We now shift the Hamiltonian's spectrum by adding a term proportional to the identity,
\begin{align*} 
H' &= H + \ii \sum_{S\in \mathcal{P}} |d_S|  \nonumber\\
   &= \sum_{S\in \mathcal{P}} |d_S| \left( \ii + \textrm{sign}(d_S) S \right) 
= \sum_{S\in \mathcal{P}} 2|d_S| P_S,
%\label{Hshift1} 
\end{align*}
which gives us a weighted sum of $k$-local projectors of the form
$P_S = \frac{1}{2} \left( \ii +\textrm{sign}(d_S) S\right)$.
We now further rescale $H'$, getting
\begin{align*} 
H'' &= \frac{1}{\sum_S 2|d_S|} H' = \sum_{S} \pi_S P_S, 
%\label{Hshift2} 
\end{align*}
another weighted sum of projectors, whose weights
\begin{align*}
\pi_S = \frac{|d_S|}{\sum_S |d_S|}\ge0
\end{align*}
now sum to 1, and thus defines a probability distribution $\pi$ on a list
of at most $k$-local Pauli operators.  The ground state of $H$ is also the
ground state of this shifted and rescaled $H''$.  Note that all of these
transformations can be carried out classically using only polynomial
resources.

Arthur now asks Merlin to send the qubits of the ground state of $H$
one-by-one.  Arthur can estimate the energy of the state $\ket{\psi}$ he
receives using the following single-qubit-measurement verification
procedure:
\begin{enumerate}
	\item Pick $S$, an at most $k$-local Pauli product, at random, according to the distribution $\pi$. 
\item Ask Merlin to send the qubits of the witness state $\ket{\psi}$ one by one.
\item On each qubit on which $S$ acts non-trivially, measure the
corresponding single-qubit Pauli operator.  Take the list of results
$x_i = \pm 1$ for $i=1,...,k$, and calculate the quantity
$r = \frac{1}{2}\left( 1 +\textrm{sign}(d_S) x_1 x_2 \dots x_k \right)$, which can take the value $0$ or $1$.

\item Accept if $r=0$.

\end{enumerate}

\if0
This works, because the expectation value of each of the $r$'s (as well as of their average) is
	\begin{align} 
		\langle r \rangle	&= \frac{1}{2} \pm \frac{1}{2}\sum_{S} w_S \bra{\psi} S \ket{\psi} 
					\nonumber\\
			&= \sum_{S} w_S \bra{\psi} P_S \ket{\psi} 
			= \bra{\psi} H'' \ket{\psi},
			\label{expectR} 
	\end{align}
equal to the expectation value of the energy of the state $\ket{\psi}$ for the Hamiltonian $H''$. To get an expectation value of the Hamiltonian $H$ in the state $\ket{\psi}$, we just need to shift and rescale it as in \eqref{rrescale}.
It remains to show that this procedure is sound, and to calculate the new completeness/soundness parameters.
\fi

In the ``yes'' case, the ground state energy is promised to be $\le E_a$, and 
the single-shot probability of obtaining the result $1$ when looking at $r$ is 
\begin{align}
	p^{\textrm{yes}}(1) &= \langle r \rangle = \bra{\psi} H'' \ket{\psi} 
	= \frac{1}{\sum_S 2|d_S|} \bra{\psi} H' \ket{\psi}\nonumber \\
	&= \frac{1}{\sum_S 2|d_S|} \left(\bra{\psi} H \ket{\psi} 
			+ \sum_S |d_S|\right) \label{expectHfromR}\\
	&\leq \frac{E_a}{\sum_S 2|d_S|} + \frac{1}{2}
\nonumber,
\end{align}
when Merlin sends us a good witness $\ket{\psi}$, whose energy (for the Hamiltonian $H$) is $\le E_a$. We accept if we measure $0$, so the acceptance probability in this case is at least
\begin{align}
	p^{\mathrm{yes}}_{\textrm{acc}} = 1- p^{\textrm{yes}}(1) \geq \frac{1}{2} - \frac{E_a}{\sum_S 2|d_S|}.
	\label{yesLH}
\end{align}

On the other hand, in the ``no'' case, 
all states have energy that is guaranteed to be $\ge E_b$. With this in mind, we can just 
look at Eq.~\eqref{expectHfromR} and find a lower bound in terms of 
$E_b$.
The single-shot probability of measuring the result 1 in our procedure is bounded from below by
\begin{align*}
	p^{\textrm{no}}(1) &\geq \frac{E_b}{\sum_S 2|d_S|} + \frac{1}{2}.
\end{align*}
Thus, Arthur will measure $r=0$ and accept a false proof in this case with probability
\begin{align}
	p^{\textrm{no}}_{\textrm{acc}} = 1- p^{\textrm{no}}(1) &\leq \frac{1}{2} - \frac{E_b}{\sum_S 2|d_S|}.
	\label{noLH}
\end{align}

Putting together Eq.~\eqref{yesLH} and 
Eq.~\eqref{noLH} we obtain a gap in the acceptance probabilities between the ``yes'' and ``no'' cases:
\begin{align*}
p^{\textrm{yes}}_{\textrm{acc}} - p^{\textrm{no}}_{\textrm{acc}} \geq \frac{E_b-E_a}{\sum_S 2|d_S|}.
\end{align*}
As the original Hamiltonian $H$ contains $M=\mathrm{poly}(|x|)$ terms
$H_m$ of bounded strength, and each $H_m$ has contributions from
$4^k=\mathrm{poly}(|x|)$ Pauli products [as long as $k=O(\log(|x|))$], we
have that $\sum_S|d_S|\le\mathrm{poly}(|x|)$ from above, and thus 
have obtained a $1/\textrm{poly}(|x|)$ completeness-soundness gap for our
1-qubit measurement procedure.

This scheme can be used to construct a QMA verification using single qubit
measurements only: Given an instance of a QMA problem, Arthur rewrites the
proof as a Local Hamiltonian $H$ and asks Merlin for the ground state of
$H$. Using the outlined procedure, Arthur accepts with $p^\mathrm{yes}_{\textrm{acc}}$
and $p^\mathrm{no}_{\textrm{acc}}$, respectively, putting the problem in
$\textrm{QMA}\left(p^\mathrm{yes}_{\textrm{acc}},p^\mathrm{no}_{\textrm{acc}}\right)=\textrm{QMA}$. Alternatively,
this can be seen as a scheme to estimate the ground state energy with
single qubit measurements only, where multiple copies of the ground state
can be used to obtain a more accurate estimate.

Note that the construction presented in this Section does not work for QMA$_1$.
For problems in this complexity class one needs to be able to accept
perfect
witnesses/proofs with 100\% probability (perfect completeness). However,
our (single qubit measurements only) procedure for estimating the
expectation value of a $k$-qubit operator is inherently probabilistic for
any $k$-local operator that is not a a product of Pauli operators; thus,
there is a finite probability that we will reject even a perfect witness.

%%%%%%%%%%%%%%%%%%%%%%%%%%%%%%%%%%%%%%%%%%%%%%%%%%%%%%%%%%%%%%%%%%%%%%%%%%%%%%%%%%%%%%%%%%

\acknowledgments
We thank Harumichi Nishimura and Matthew McKague for valuable discussions,
and Joseph Fitzsimons for pointing out Ref.~\cite{JiVerification}.  TM is
supported by the Grant-in-Aid for Scientific Research on Innovative Areas
No.15H00850 of MEXT Japan, and the Grant-in-Aid for Young Scientists (B)
No.26730003 of JSPS. DN's research leading to these results has received funding from the People Programme (Marie Curie Actions) EU's 7th Framework Programme under REA grant agreement No. 609427. This research has been further co-funded by the Slovak Academy of Sciences.
DN was also supported by the grant QIMABOS APVV-0808-12, and thanks Frank Verstraete's group in Vienna, where part of this work was done. NS is supported by the Alexander von Humboldt-Stiftung and the European Union (projects QALGO and WASCOSYS). DN and NS thank the Simons Institute for the Theory of Computing in Berkeley, where parts of this
work were carried out, for their hospitality.

%%%%%%%%%%%%%%%%%%%%%%%%%%%%%%%%%%%%%%%%
%\bibliographystyle{apsrev4-1}	% (uses file "apsrev4-1.bst")
\bibliography{QMAsimple}

\end{document}